\documentclass[a4paper,twoside]{article}

\usepackage{epsfig}
\usepackage{subcaption}
\usepackage{calc}
\usepackage{amssymb} 
\usepackage{amstext} 
\usepackage{amsmath} 
\usepackage{amsthm}   
\usepackage{multicol}
\usepackage{pslatex} 
\usepackage{apalike}
\usepackage{SCITEPRESS}     

\usepackage{algorithm} 
\usepackage{algorithmicx}
\usepackage{algpseudocode}
\algrenewcommand\algorithmicrequire{\textbf{Input:}}
\algrenewcommand\algorithmicensure{\textbf{Output:}}

\usepackage{pdflscape}
\usepackage[pass]{geometry}

\begin{document}

\title{Visually Improved Erosion Algorithm for the Procedural Generation of Tile-Based Terrain}

\author{\authorname{Fong Yuan Lim, Yu Wei Tan and Anand Bhojan \orcidAuthor{0000-0001-8105-1739}}
\affiliation{Department of Computer Science, National University of Singapore, Singapore}
\email{e0169576@u.nus.edu, yuwei@u.nus.edu), banand@comp.nus.edu.sg}
}


\abstract{Procedural terrain generation is the process of generating a digital representation of terrain using a computer program or procedure, with little to no human guidance. This paper proposes a procedural terrain generation algorithm based on a graph representation of fluvial erosion that offers several novel improvements over existing algorithms. Namely, the use of a height constraint map with two types of locally defined constraint strengths; the ability to specify a realistic erosion strength via level of rainfall; and the ability to carve realistic gorges. These novelties allow it to generate more varied and realistic terrain by integrating additional parameters and simulation processes, while being faster and offering more flexibility and ease of use to terrain designers due to the nature and intuitiveness of these new parameters and processes. This paper additionally reviews some common metrics used to evaluate terrain generators, and suggests a completely new one that contributes to a more holistic evaluation.}

\onecolumn \maketitle \normalsize \setcounter{footnote}{0} \vfill

\section{\uppercase{Introduction}}

\begin{figure*}[ht]
  \centering
  \includegraphics[width=\textwidth]{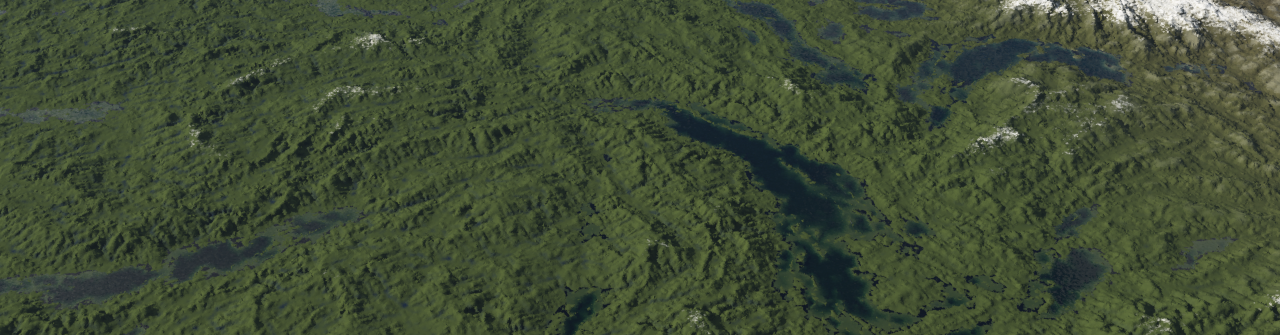}
  \caption{Sample output of the algorithm, rendered in Terragen 4. \label{fig:teaser}}
\end{figure*}

Terrain graphics are integral to many applications in the real world, primarily in the entertainment industry such as games and movies, but also holding value in various other industries such as education, training and engineering. Capturing and rendering real terrain has several drawbacks, such as requiring a large amount of storage space, losing detail at high zoom levels, and covering a finite area.

Procedural terrain generation (PTG) is the generation of terrain by an automated process with limited to no human input, which overcomes the limitations of using only hard terrain data. Due to its automated nature, it is a scalable method for generating large amounts of content with minimal costs in time and human labour. There are professional PTG tools such as \textit{Terragen}, \textit{World Creator} and \textit{E-on Vue} in the industry today, and many successful movies and games have utilized PTG in their production as well, such as \textit{Pirates of the Caribbean: Dead Man's Chest}, \textit{The Golden Compass}, \textit{Far Cry 5}, and \textit{Minecraft}.

Despite the benefits of PTG, there remain limitations and room for further improvement in various aspects, such as efficiency, variety and realism.
This paper proposes an erosion-based PTG algorithm that offers improvements in several visual and technical aspects over current solutions. A teaser output of the algorithm is shown in Figure \ref{fig:teaser}. The contributions of this work are as follows:
\begin{enumerate}
\item A new way to specify terrain constraints in terms of a constraint map with local value and gradient constraint strengths empowers the terrain designer with greater flexibility, accuracy and intuition in controlling the output, along with making the simulation converge faster.
\item The additional moisture parameter is a powerful dictator of how much erosion should occur at a particular region of terrain, and thus provides a high degree of controllability to the terrain designer over the distribution of flat vs mountainous areas, as well as greater variety and realism.
\item The algorithm's ability to generate convincing gorges improves the realism of the output terrain.
\item Proposal of a set of metrics, including an original metric, to evaluate terrain generators with. Modern terrain generators and the new proposed generator are compared using these metrics.
\end{enumerate}

The rest of this paper is structured as follows:
Section \ref{ch:relatedWork} briefly discusses related work.
Section \ref{ch:design} explains the proposed algorithm.
Section \ref{ch:evaluation} analyzes the algorithm and compares it to existing solutions.
Section \ref{ch:conclusion} concludes this paper and highlights some potential areas for future research.

\section{\uppercase{Related Work}}
\label{ch:relatedWork}
In this section, we discuss the current PCG algorithms in four categories, Stochastic Algorithms, Erosion Simulation Algorithms, Graph-based Algorithms and Machine 
Learning. 
\subsection{Stochastic Algorithms}
The most common method to automatically generate terrain by far has been with fractal noise. This method originated in \cite{Perlin85}, and is explained in \cite{PCGGamesBook} chapters 4.2-3, as well as various online tutorials.


This approach has been further improved in \cite{Perlin02}. An alternative noise generation algorithm called simplex noise that improved upon the shortcomings of Perlin noise was also presented in \cite{Perlin01}.

The diamond-square algorithm \cite{Miller86} is an alternative approach to this method that produces a similar result, albeit with more axis-aligned artifacts.


\cite{Belhadj07} constrains the diamond square algorithm to be of a certain height at user-specified points, thus demonstrating that the method is controllable. This method was successfully used to increase the resolution of satellite DEMs.

\cite{Bangay17} attempts to avoid some common pitfalls of other methods, such as visible seams and repetitive elements, by working with gradient maps instead of height-maps. A variant of Wang tiling is used to generate the gradient maps, and the height-map is reconstructed from there by solving a related Poisson equation. The designer has the freedom to specify what the Wang tiles are, which creates overlap with an example-based methodology, but the algorithm otherwise pieces the Wang tiles together stochastically with no designer input. The resulting terrain has a messy, crumpled appearance, which is somewhat unrealistic especially at lower elevations.

\cite{Gasch20} presents a noise-based method that allows procedural terrains creation with elevation constraints. 

\subsection{Erosion Simulation Algorithms}
The effects of rivers and erosion have been recognized as the most important contributing factor to the shape of terrain as early as \cite{Kelley88} and \cite{Musgrave89} respectively. Erosion simulation algorithms excel at producing realistic terrain, particularly the generation of creases on hillsides that strongly imply water erosion. However, their main drawback is a slow generation speed.

\cite{Benes02} elaborates on hydraulic erosion and its basis in the real world. \cite{Benes06-1} and \cite{Benes06-2} base hydraulic erosion on the Navier-Stokes equations that describe viscous fluid dynamics in real world physics. \cite{Neidhold05} augments the standard algorithm with interactive terrain manipulation tools such as painting water or water sources directly onto the terrain as the simulation is running.

\cite{Musgrave89} is an early paper that simulates both thermal weathering and hydraulic erosion. For hydraulic erosion, the terrain height as well as amount of water and sediment are tracked for each pixel. As water flows out of a pixel into neighbouring pixels, a proportionate amount of sediment is transported as well. The ground is eroded or sediment is deposited depending on the sediment carrying capacity of the water. A fixed amount of water is added to every pixel at regular intervals to simulate rain. \cite{St'ava08} adopts the same approach for hydraulic erosion, and adds that water penetrates into the ground and turns it into slushy regolith, which is simulated as a slow-moving liquid.

\cite{Hudak11} is able to simulate landslides by tracking the wetness of the soil. However, the soil is represented as discrete particles, which is unrealistic for large-scale terrain and computationally expensive as it involves particle collisions.

More recent papers focus on the efficiency of such algorithms. \cite{Hoang-Anh07} and \cite{Mei07} propose to run the simulation on the GPU. \cite{Vanek11} proposes to speed up hydraulic erosion simulation by using a quadtree representation for the terrain.

The standard way to simulate water, as used in the majority of the papers cited above, is to track the amount of water held by each tile, then distributing that water to its neighbouring tiles. This mechanism restricts water to flow at a maximum of one tile per simulation tick, which is highly unfeasible for simulation at geological time and space scales.

\subsection{Graph-based Algorithms}
Graph-based algorithms attempt to draw a graph - typically a river network - onto a terrain, then use this graph to fill in the terrain height. This may be done in either a stochastic or a simulation fashion. Graph-based algorithms tend to be more structured and realistic than the standard stochastic methods due to their attention to the dendritic nature of real terrain features, and they are generally faster than the standard simulation methods due to their efficient way of defining rivers and water flow.

\cite{Kelley88} defines an initial main river, then generates a complete river network by recursively generating branches along the river, calculates the height at varying points on the river based on how much erosive power that part of the river has, and finally constructs the rest of the terrain from using the heights of the points on the river. \cite{Genevaux13} adopts the same general approach, with more developed geological theories. \cite{Zhang16} also uses the same general approach, but they generate the terrain between rivers using the midpoint displacement algorithm, and their usage of an L-system to generate the river network results in distinct regions of terrain with regular hills.

\cite{Gaillard19} generates a random dendritic fractal shape that can be explicitly used as mountain ridges. The height of the terrain at any point is proportionate to its distance from the closest branch of the fractal. The fractal can be controlled to take on a particular shape.

\cite{Cordonnier16} implements erosion simulation on a graph-based terrain. The underlying philosophy of the algorithm is that terrain is shaped by two processes: tectonic uplift and river erosion. Uplift raises the terrain, while erosion lowers the terrain. The simulation runs until the terrain reaches a stable state where the uplift balances out the erosion at every pixel.

Every pixel of the height-map is given a point with a random position within it. Each point is also associated a fixed drainage area - representing the area of locality from which it collects water, and thus how much water has to be drained from it. Finally, each pixel also keeps track of an uplift value - how much the terrain (rock) at that point is raised every tick.

Within a simulation tick, uplift is applied every pixel, then erosion. For the erosion step, every pixel first identifies its neighbour with the lowest height. This represents the neighbour that all its collected water is going to flow to. If all its neighbours are higher than it, it aborts the erosion step. Once every pixel determines its lowest neighbour, the total drainage $A$ of each pixel is calculated. The total drainage is the total sum of the current pixel's drainage area and the drainage areas of everything upstream of it. Simultaneously, the slope $s$ of each pixel is calculated to be the gradient between its own point and the point of the neighbour that it is emptying into. Finally, erosion is calculated as $ks\sqrt{A}$, where $k$ is an experimentally derived erosion constant. Every pixel's terrain height is reduced by this erosion value for this tick.




\subsection{Machine Learning}
Machine learning (ML) has in recent years found many applications throughout the field of information technology and beyond, and terrain generation is no exception. When one regards terrain height-maps as images, it is clear that existing ML techniques are highly compatible with height-map generation, given a large enough database of existing terrain samples to learn from.
\cite{Summerville18} is a survey on ML techniques being used for PCG for functional elements in games.

\cite{Yeu06} is an early attempt at generating terrain using ML techniques - in this case, an extreme learning machine (ELM). \cite{Guerin17} proposes to use a Conditional Generative Adversarial Network (cGAN) that has been trained on real terrain in order to generate new terrain. \cite{Wulff18} uses a Deep cGAN to generate alpine height-maps. \cite{Spick19} uses a GAN to generate both a height-map and a corresponding satellite image.

ML techniques have the capability to produce high-quality terrain output with fast speed and simple inputs. However, they can only do so with the help of many auxiliary resources: First of all, there needs to be a database of realistic and properly formatted terrain samples to learn from. Then, the algorithm itself has to be engineered to learn from these samples effectively. Finally, a terrain designer has to specify where major terrain features should actually be. Without any of these components, the quality of terrain output is heavily compromised.  
\section{\uppercase{Design}}
\label{ch:design}
In this section, we will describe our terrain generation algorithm and we evaluate it for its performance and quality in next section. 
\subsection{Background \& Overview}
Our algorithm represents the {\em terrain} with a layered height-map \cite{Benes01}. However, what is tracked per tile is not layers of material, but more abstract quantities such as hardness and moisture that are discussed further in this section. This terrain representation was chosen to provide compatibility with other software and the opportunity for interactive manipulation. The abstract quantities are core to the simulation itself, but may also prove useful in later stages of the terrain generation pipeline, such as moisture determining where vegetation grows.


Each heightmap tile tracks the following persistent quantities:
\begin{itemize}
\item \textbf{Node offset (x,y)}: Each tile is treated as a point position (a node), and these coordinates represent the node's position offset from the bottom left corner of the tile, ranging from 0 to 1 each. Variance in this offset minimises axis-aligned artifacts in the terrain, by varying the distance and thus gradient with respect to the neighbouring tiles.
\item \textbf{Land height}: The height of the terrain in the tile.
\item \textbf{Water height}: The amount of water in the tile. This essentially indicates that the tile is part of a water body, and it should not undergo erosion. The total height of a tile refers to the sum of its land height and water height.
\item \textbf{Height constraint and constraint strengths}: The ideal land height of the terrain, and how strongly to bind the terrain to this ideal height.
\item \textbf{Moisture}: How much rain falls on the tile.
\end{itemize}
There are additional quantities that each tile keeps track of, but they are only temporary and will be reset at the end of the simulation phase or simulation tick. They will be brought up and explained in the later subsections as they become relevant to the simulation.

On algorithm initialization:
\begin{itemize}
\item \textbf{Node offset} $\in [0,1]^2$: Each of $x$ and $y$ are independently set to a uniformly distributed random value between 0 and 1.
\item \textbf{Height constraint}: Specifiable.
\item \textbf{Height value constraint strength} $\in [0,1]$: Constant low value, such as 0.02. Alternatively, the same as the height constraint, re-mapped to be between 0 and 1.
\item \textbf{Height gradient constraint strength} $\in [0,1]$: Specifiable. Generally, a uniform low value can be used for a height constraint map with low detail, and a uniform high value can be used for a height constraint map with high detail.
\item \textbf{Moisture} $\geq 0$: Specifiable. It could be set to random Perlin noise, or a uniform high value.
\item \textbf{Land height}: Initialized to be the same as the height constraint.
\item \textbf{Water} $\geq 0$: Any tile below sea level is filled with water until it becomes sea level.
\end{itemize}
The terrain designer ideally loads an initial height-map as the height constraint map. To prevent unnatural behaviour when simulating the rivers, small random noise is added to the constraint map.

The simulation phase comes after the initialization phase. On each simulation tick, the set of processes are executed on the terrain are shown in Algorithm 1.

\begin{algorithm} [H]
    \label{AlgoSim}
    \caption{Simulation Algorithm}
    \begin{algorithmic}
	    \For {Each Simulation Tick}
            \begin{enumerate}
    		\item For each tile, determine a neighbouring tile (one of the four tiles adjacent to it, i.e. its Von Neumann neighbourhood) to drain all its water to, if any.
			\item Calculate the total amount of water flowing into each tile (total drainage), which is the sum of how much rain it catches, and how much water it receives from its tributary neighbours.
			\item Connect local minima together by carving gorges between them.
			\item Erode each tile based on how much total drainage it has and its slope.
			\item Apply the height constraints.
			\item Reset the sea level.
            \end{enumerate} 
	    \EndFor
	\end{algorithmic}
\end{algorithm}

Each step is elaborated in the further subsections below.


\subsection{Drainage Direction Calculation}
It makes physical sense that the water in a tile will be drained to some neighbour of a lower total height than the tile. If more than one such neighbour exists, we wish to drain to at most one neighbour only, in order to exploit the tree structure that emerges. We choose this neighbour as the one which creates the steepest gradient with the current tile, calculated as their total height difference divided by the horizontal distance between the two tiles' nodes. If no such neighbour exists, then this tile does not drain to any node, and is a local minimum on the terrain.

Each tile also tracks which neighbours drain to it. These neighbours are called tributaries.
To refer to more distant tiles along the same drainage network, the terms upstream tiles and downstream tiles are used.

The local minima are recorded down, to be utilised later for erosion. Minima that have water on them are discarded, as erosion should not occur on them.

\subsection{Drainage Calculation}
Each tile has a local drainage and a total drainage value. The local drainage indicates how much rain is collected by that tile, which is simply the moisture parameter of that tile. The total drainage is the total amount of water collected by each tile from all sources, which is the sum of the local drainage, and the total drainage from its tributary neighbours.
Naturally, the total drainage of a leaf node is simply its local drainage, as it has no tributaries.

The contribution from the tributaries is reduced by a constant factor $k_d \in [0,1]$ at each tile, with a cumulative effect on tiles further upstream. Physically, this represents water loss over the course of a river, due to various processes such as evaporation and seepage into the ground. Mechanically, this helps to prevent too much water and thus too much erosion happening downstream. $k_d$ is usually set to $0.68$.

The total drainage is calculated by recursively calculating the total drainage of the tile's tributaries, multiplying it by $k_d$, then adding on the local drainage. Due to the tree structure of the drainage network, the total drainage of all tiles may be accurately calculated in linear time with respect to the number of tiles in the simulation.

\subsection{Minima Connection / Gorge Carving}
Simply eroding based on the drainage basin formed is not good enough, as the drainage basin quickly becomes static, and forms very cellular regions on the terrain, separated by mountain ranges of unnatural straightness and uniformity in height.
To counteract this, drainage basins need a mechanism to connect to one another in some way. However, the difficulty in doing so is maintaining the connection over a path of greater resistance, in the sense that water has to drain uphill, antagonistic to the terrain height and rate of height increase.

For each local minimum $M$, a depth-first search is performed to iterate through all its leaf nodes. The leaf node with the lowest total height is selected. Its neighbour in the opposite direction of the current neighbour it is draining to is also noted. By running downslope from both of these slopes, a complete path from the $M$ to an adjacent local minimum $M'$ can be derived. Along this path, the ideal height of each node $n$ is calculated as the linearly interpolated height of the two local minima, with respect to the distance along the path. The height of $n$ is then updated to be a linear weighted average of its current height and its ideal height, if the current height is above the ideal height. The weight $w \in [0,1]$ is a function of the total drainage $D(M)$ of $M$. $w$ should be non-decreasing with respect to $D(M)$, be 0 when there is no drainage (so there is no carving), and approach 1 when there is high drainage (so carving is at full strength). The function used in the experiments is
\begin{equation}
w = \max\{k_g \sqrt{D}, 1\}
\end{equation}
where the constant $k_g$ is set to 0.1.

To preserve the gorge, the height gradient constraint strength of each node along the path is also weakened, by a factor of $1-w$.

\subsection{Fluvial Erosion}
All tiles that drain to another tile are subject to fluvial erosion. The decrease in height due to fluvial erosion is given by the stream equation:
\begin{equation}
\Delta h = k_e D^n s^m
\end{equation}
where
\begin{enumerate}
\item $k_e$ is the erosion constant, and usually set to $0.5$.
\item $D$ is the total drainage on that tile.
\item $s$ is the slope of that tile, calculated as the gradient from this tile to the tile it is draining to.
\item $n$ and $m$ are arbitrary constants. Only their ratio seems to affect the actual look of the terrain, and the ratio $n/m = 1/2$ seems most ideal. We thus set $n = 1$ and $m = 2$.
\end{enumerate}
The decrease in height is capped at the difference in height between the current tile and the tile it is draining to. This is to prevent an oscillating feedback loop between two adjacent tiles alternately draining into each other, and pulling the higher one below the lower one, creating an infinitely sinking hole.

\subsection{Height Constraints}
Height constraints are applied to the terrain in order to let it maintain its shape and altitude amidst the erosion processes. There are two types of constraints employed: the value constraint and the gradient constraint.

The value constraint simply attempts to pull the height of the terrain back to its ideal height, as specified in the height constraint map. The height of each tile is updated to be a linear weighted average of its current height and its ideal height, with the weight being the strength of the value constraint.

The role of the value constraint is mainly to ensure that the whole terrain does not continuously sink to a lower and lower height, due to the erosion processes. At the same time, if it is set too high, then the terrain is too resistant to change and the erosion processes will have no effect on the terrain. Thus, it is ideal to set the value constraint to a low uniform value. In our experiments, we set it to $0.02$. Even though this constraint is conceptually able to pull terrain both up and down, in practice, it only pulls the terrain up, as all the other processes only pull the terrain down.

The gradient constraint attempts to preserve the local height relationship between adjacent pixels.
Given a tile $t$, each of its neighbours $n$ asserts the most desired height $h'_{t,n}$ of $t$ as $h_n + (c_t - c_n)$, where $h_x$ and $c_x$ are the current height and constraint height of tile $x$ respectively. The ideal height $h'_t$ for $t$ is simply the average of these desired heights from all the neighbours. The height of $t$ is then updated to be a linear weighted average of its current height and its ideal height, with the weight being the strength of the gradient constraint.

\begin{figure}[ht]
    \centering
    \renewcommand{\arraystretch}{0.5}
    \renewcommand{\tabcolsep}{1pt}
	\begin{tabular}{c c c}
        Initial & High Strength & Low Strength\\
		
		\includegraphics[width=.33\linewidth]{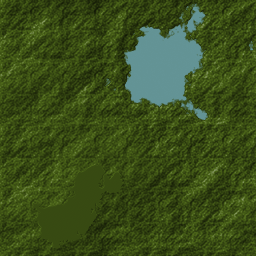} &
		\includegraphics[width=.33\linewidth]{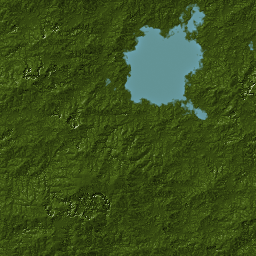} &
		\includegraphics[width=.33\linewidth]{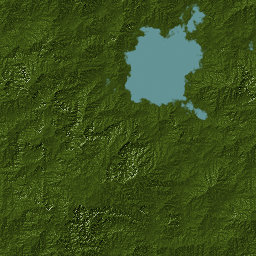} \\
		
		\includegraphics[width=.33\linewidth]{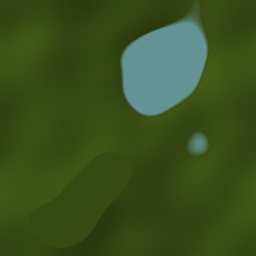} &
		\includegraphics[width=.33\linewidth]{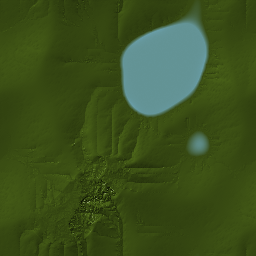} &
		\includegraphics[width=.33\linewidth]{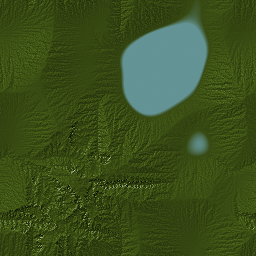} \\
    \end{tabular}
    \caption{Comparison of a high gradient constraint strength (0.8) with a low gradient constraint strength (0.02) on two constraint maps, filled with 3 and 8 octaves of fractal Perlin noise respectively. 100 iterations. Value constraint strength of 0.02. Uniform moisture of 1. \label{fig:gradient_strength}}
\end{figure}

\begin{figure}[ht]
    \centering
	\renewcommand{\arraystretch}{0.5}
    \renewcommand{\tabcolsep}{1pt}
	\begin{tabular}{c c c}
        Initial & 100 iterations\\
		\includegraphics[width=.5\linewidth]{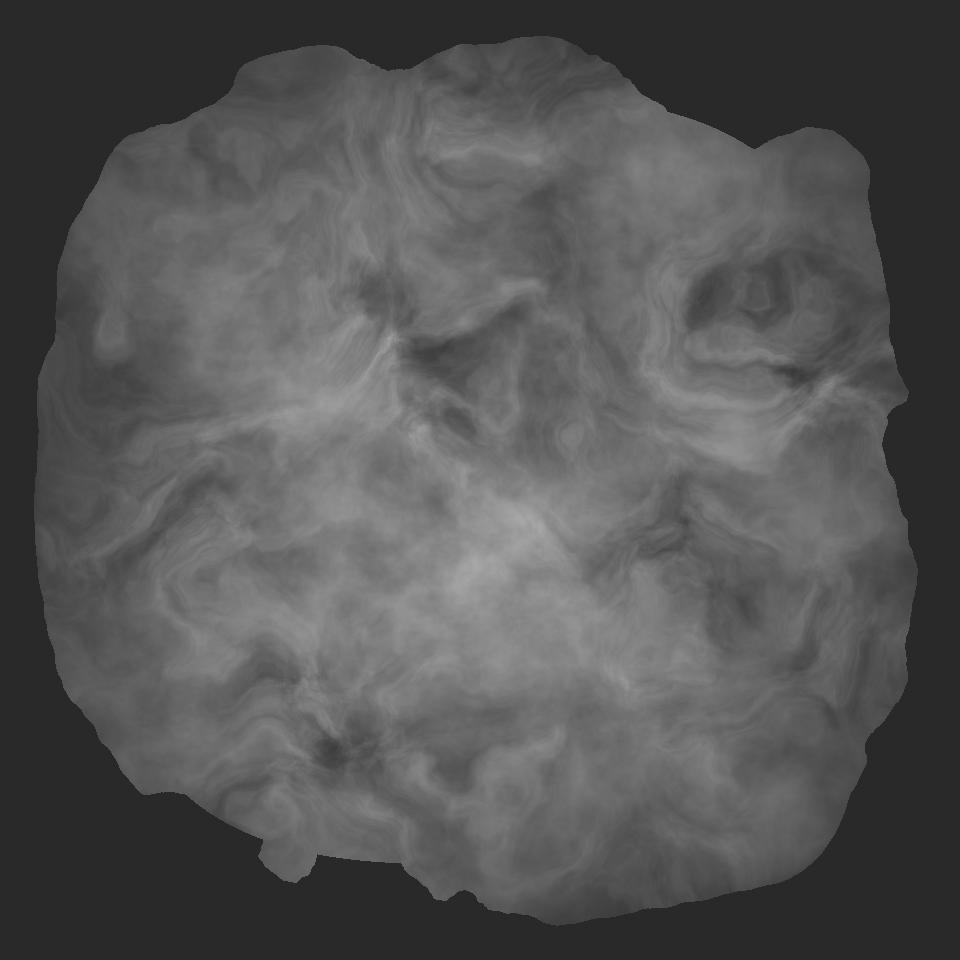} &
		\includegraphics[width=.5\linewidth]{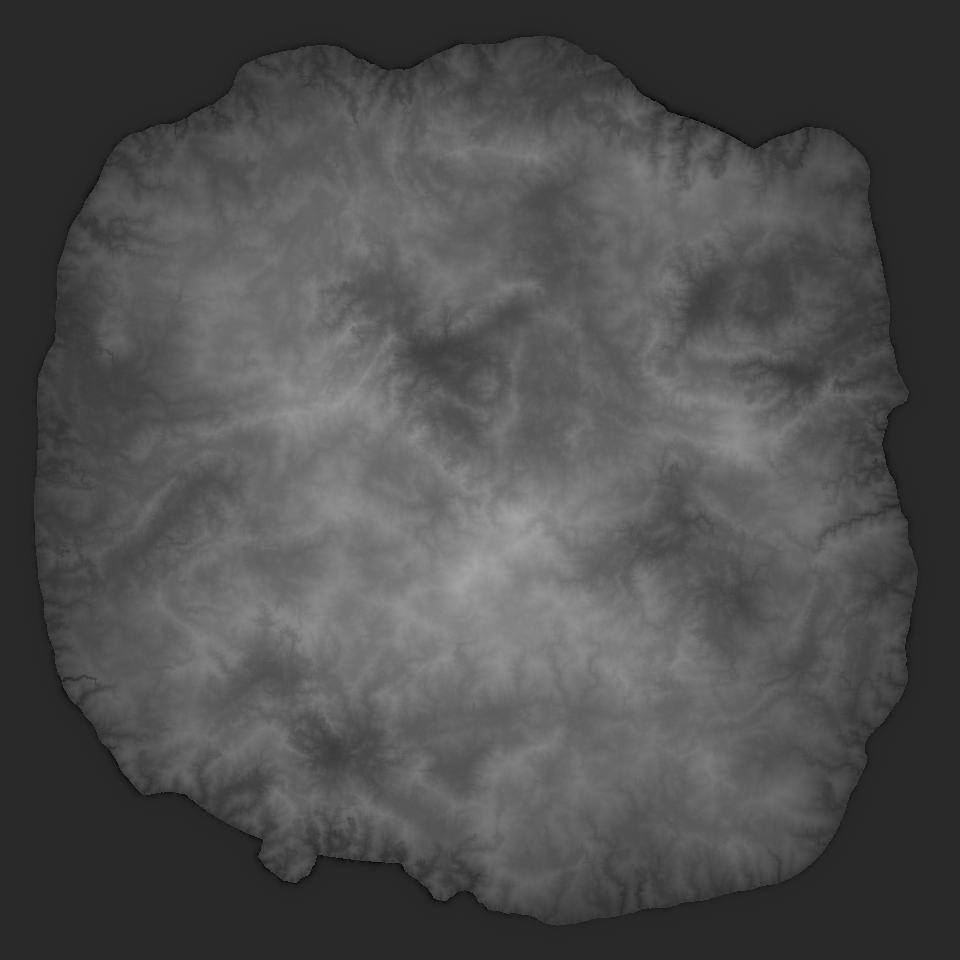}
    \end{tabular}
	\includegraphics[width=\linewidth]{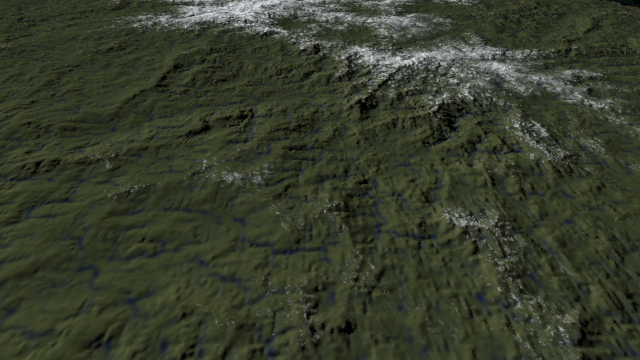}
    \caption{Sample output heightmap of the algorithm, 960x960, after 100 iterations. Value constraint strength of 0.02. Gradient constraint strength of 0.8. Uniform moisture of 1. \label{fig:sample}}
\end{figure}

A high gradient constraint strength preserves details that are present in the constraint map, and thus should be used if the constraint map has desirable local features. Conversely, a low gradient constraint strength gives the erosion processes more freedom to shape the terrain, and would be ideal to use in regions where the constraint map has few or undesirable local features. The terrain designer can exploit this by specifying a high strength over regions with high detail, and low strength over regions that they want to erode and add detail to, such as cliffs and smooth surfaces. In our experiments (Figure \ref{fig:gradient_strength} and \ref{fig:sample}), we consider $0.8$ a good value for high strength and $0.02$ for low strength.

\subsection{Sea Level}
For every tile, if its height is below sea level, the amount of water it has is set so that its height is sea level. Otherwise, the tile will have no water.

This is a simple yet necessary step in the simulation. A tile's height may be reduced not only from the erosion processes (which would never erode below the local minimum height), but also from satisfying the height gradient constraint, which could push the height below sea level. Without adding water to compensate, the erosion processes will reinforce this dip in height and allow further dips to accumulate.

\section{\uppercase{Results}}
\label{ch:evaluation}
This section presents the output of the proposed algorithm and compares it to existing cutting-edge algorithms for performance and quality.
\subsection{Performance}
\begin{figure}[ht]
    \centering
    \renewcommand{\arraystretch}{0.5}
    \renewcommand{\tabcolsep}{1pt}
	\begin{tabular}{c c c}
        Cordonnier's algorithm & Our algorithm \\
		
		\includegraphics[width=.5\linewidth]{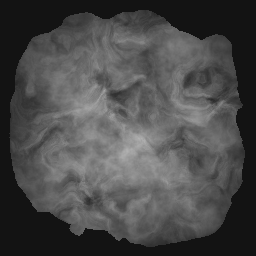} &
		\includegraphics[width=.5\linewidth]{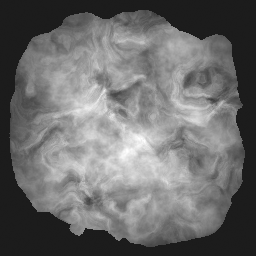} \\
		
		\includegraphics[width=.5\linewidth]{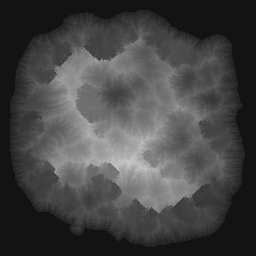} &
		\includegraphics[width=.5\linewidth]{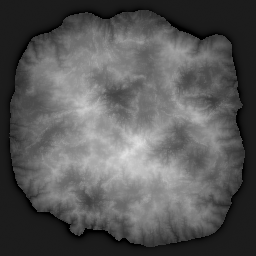} \\
		
		\includegraphics[width=.5\linewidth]{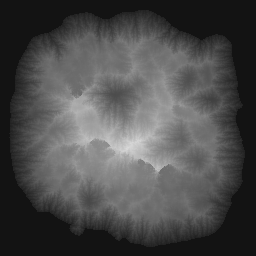} &
		\includegraphics[width=.5\linewidth]{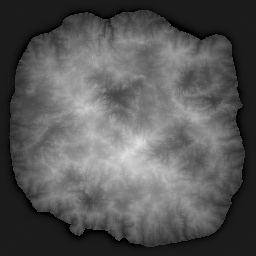} \\
		
		\includegraphics[width=.5\linewidth]{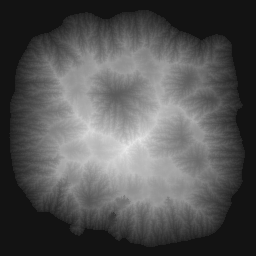} &
		\includegraphics[width=.5\linewidth]{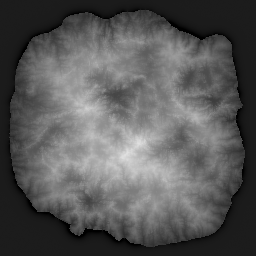} \\
    \end{tabular}
    \caption{Side-by-side comparison of the output of Cordonnier's \cite{Cordonnier16} and our algorithms. From top to bottom: State of the terrain at 0, 50, 100 and 150 iterations. \label{fig:algo_convergence}}
\end{figure}

Table \ref{tab:algo_time} below lists some timing results for our algorithm and the graph-based erosion algorithm from \cite{Cordonnier16} ("Cordonnier's algorithm"). The input is as shown in Figure \ref{fig:algo_convergence}. The algorithms were implemented in C++ and tested on a machine running 64-bit Windows 7 with an Intel\textregistered Core\texttrademark i5-4690 processor clocked at 3.50 GHz, and 8 GB of RAM. The results are the average of 10 measurements.

As shown, our algorithm runs only marginally slower than Cordonnier's algorithm. Furthermore, when comparing the output of the two algorithms (see Figure \ref{fig:algo_convergence} below), our algorithm converges to a realistic output in a fewer number of iterations - achieving a realistic output in 50 iterations compared to 150, and thus takes less time overall.
\begin{table}[ht]
	\begin{center}
	\caption{Time taken in seconds (s) for Cordonnier's \cite{Cordonnier16} and our algorithms to generate a $n\times n$ square of terrain with 100 iterations. \label{tab:algo_time}}
	\begin{tabular}{ |c||c|c| }
		\hline
		$n$ & Cordonnier's algorithm & Our algorithm \\ \hline
		 256 &   7.3 &   7.5 \\
		 512 &  29.4 &  31.3 \\
		1024 & 124.8 & 115.6 \\ \hline
	\end{tabular}
	\end{center}
\end{table}

Similarly demonstrated in Figure \ref{fig:algo_convergence} is the increased realism our algorithm provides even at later iterations. Cordonnier's algorithm generates highly cellular drainage basins, separated by very chiselled and uniform mountain ranges. Both are unrealistic features, and furthermore destroy any visually interesting details and directional biases in the uplift map. The basins merge and grow bigger and stronger as the simulation stabilizes, exacerbating this effect. Our algorithm resolves this by directly reducing the height over the drainage path, in order to produce much more pronounced gorges that influence the shape of the drainage basins more drastically. In addition, our replacement of the uplift map with a constraint map system allows us to respect aforementioned visually interesting properties in the initial terrain.

The tile-based representation of our terrain allows for easy adaptation and modification for further techniques. For example, the concept of a talus angle - the maximum angle the terrain can make with the horizontal before material starts to fall down the slope - may be implemented by asserting that terrain is not allowed to rise past the lost maximum height allowable by its neighbours and the talus angle.

\subsection{Metrics and Comparison}
A list of useful evaluation metrics for generators are provided at \cite{Cantero14}, some of which are shared by \cite{Doran10}. 
Some of them relevant to our work are:
\begin{itemize}
\item \textbf{Efficiency}: How fast the generator is expected to run before outputting a final terrain result. While conventional time complexity analysis may be applied, most algorithms already run at an optimal $O(A)$ time where $A$ is the area of the generated terrain. (Algorithms that achieve even lower complexity typically use resolution-independent representations of terrain.) It is instead more useful to measure the real time taken to generate terrain directly.
\item \textbf{Interactivity}: How easily can a terrain designer affect the output of the generator. This may be divided into: \textbf{Usability}: How intuitive the controls are, opposed to abstract or technical. And \textbf{Controllability}: How much power the designer has in controlling the final terrain, such as the placement and appearance of various terrain features.
\item \textbf{Variety}: How much expressive power the generator has, or how varied are the features that it can generate. A greater variety generates more interest, aesthetically and mechanically, for the end user, motivating them to explore and experiment with the terrain.
\item \textbf{Realism}: How physically plausible or sensible is the final terrain. For example, floating islands or tall thin spikes on the ground would be regarded as unrealistic in a non-fantasy setting. Realism lends to believability and immersion in the terrain.
\end{itemize}

Additionally, we believe that another metric is important for our purposes:
\begin{itemize}
\item \textbf{Independence}: How little resources are \textit{required} to set up the generator and to run the generator to produce quality output.
\end{itemize}
These aforementioned resources may be in terms of legal, financial, time and labour. A common resource is a dedicated terrain designer, which resource-constrained teams may not have (the money to hire one; the time opportunity cost for a team member to take on the role themselves; the expertise needed to control the generator etc.). While the \textit{ability} to involve a designer is always beneficial (and factored in with the Interactivity metric above), there are scenarios where it is optimal to let the algorithm make its own design decisions, such as a random map generator shipped with a strategy game, e.g. The \textit{Sid Meier's Civilization} or \textit{Heroes of Might and Magic} series, or an endless map generator in a flight simulator or exploratory or sandbox RPG, e.g. \textit{Minecraft}. While any algorithm is arguably independent by simply inputting randomly generated parameter values, the generated result may not always meet the standards of the other metrics above, and thus should be discounted. 

Some standard and recent algorithms have already been evaluated in Section \ref{ch:relatedWork}. In this section, we compare our algorithm to two other modern competitive algorithms based on the aforementioned metrics; namely, the graph-based fluvial erosion algorithm as described in \cite{Cordonnier16}, and the ML algorithm with a cGAN as described in \cite{Guerin17}.

Our algorithm improves upon \cite{Cordonnier16} in all aspects. Instead of specifying a simple uplift map, we specify a much more expressive constraint map with two types of locally varying constraint strengths. This gives the terrain designer much more freedom and intuition in specifying terrain features, improving interactivity. Whereas in Cordonnier's algorithm, the uplift map has to be used creatively and unintuitively - such as posterising it - in order to achieve desired terrain patterns. It is also common in Cordonnier's algorithm for relatively straight and uniform mountain ridges to form, almost equidistant from water bodies and thus creating very cellular drainage basins, as these ridges demarcate the edges of the drainage basins, which tend to be simple and static.
This change from an uplift map to a constraint map also lets the simulation converge faster, improving efficiency.
The additional moisture parameter is a powerful dictator of how much erosion should occur at a particular region of terrain, thus improving controllability, variety and realism in the final terrain.
The algorithm's ability to generate convincing gorges also improves the realism of the output terrain.
Finally, the new parameters may be procedurally generated without requiring manual input, thus preserving the high level of independence in the original algorithm.

As mentioned previously in Section \ref{ch:relatedWork}, the main drawback of ML algorithms such as \cite{Guerin17} are their very poor independence, whereas our algorithm maintains a high level of independence. Furthermore, our algorithm remains competitive in interactivity, variety, and realism, and may even exceed the ML algorithm in these aspects if the latter is poorly implemented.
\section{\uppercase{Conclusion and Future Work}}
\label{ch:conclusion}

This paper has proposed an improved erosion algorithm to generate highly varied and realistic terrain, with improved efficiency, and with a competitive performance both with or without manual specification of terrain features. The algorithm achieves this by specifying a constraint map with two types of constraint strengths, which provides a stronger and more expressive specification of desired terrain features. The algorithm also employs a moisture map, which dictates how much erosion happens at any given region. Finally, the algorithm is able to generate realistic gorges that cut into mountains, creating dramatic and realistic features. Overall, our algorithm works very well and it is practically feasible to integrate with the design process.

\subsection{Limitations}
The biggest limitation of this algorithm, and indeed of PTG algorithms in general, is a set of objective metrics with which to validate the quality of generators and their generated terrain. This is a difficult problem to address, as terrain is visual by its very nature, and aesthetic quality is hard to formalize. Nevertheless, visual comparison to real terrain height-maps suggest that our algorithm is at least close to being realistic.

The adaptation to a tile-based representation introduces axis-aligned tendencies in the terrain. While this is mitigated somewhat by the random tile offsets and small perturbations in height, perhaps this could be further improved by having several nodes per tile, instead of just one.

The use of a constraint map instead of an uplift map arguably detracts from the actual realism, as do several other aspects of the algorithm - such as the formula to convert the total drainage into an abstract strength for gorge carving, which were artificially formulated without justification from the real world.

\subsection{Potential Areas for Future Research}
In the future, we would like to explore various improvements to the currently proposed algorithm.

One such improvement is a way to simulate a realistic moisture distribution, based on the presence of mountains and water bodies, so that features such as rain shadows are automatically simulated.

Our algorithm accounts only for fluvial erosion, and not other forms of erosion that may be more prominent in various regions of the world. For example, glacier erosion may be more prominent in cold areas, while wind erosion may be more prominent in hot, dry areas. We have started some preliminary studies on these new areas and we will explore them in detail. 
\begin{landscape}
\begin{table}[p]
\begin{center}
\caption{Evaluation of relevant terrain generators based on established metrics. \label{tab:evaluation}}
\begin{scriptsize}
\begin{tabular*}{\linewidth}{ |p{0.06\linewidth}||p{0.14\linewidth}|p{0.24\linewidth}|p{0.19\linewidth}|p{0.13\linewidth}|p{0.14\linewidth}| }
	\hline &
	\multicolumn{3}{c|}{Generator metrics} & \multicolumn{2}{c|}{Output metrics} \\
	
	\hline &
	Efficiency &
	Interactivity &
	Independence &
	Variety &
	Realism \\ \hline\hline
	
		Stochastic Algorithms \cite{Perlin01,Belhadj07} &
	
	\textbf{Very fast}. Milliseconds. &
	
	\textbf{Poor controllability}, as generation is mostly left to random processes with no human input. However, recent works \cite{Belhadj07} allow some degree of control by constraining the terrain height according to designer specifications or by fixing river paths.
	\textbf{High usability}. The controls are in the form of points and curves, which are intuitive for the designer. &
	
	\textbf{Very high independence}. The algorithms defer most work to stochastic sources. &
		
	\textbf{Poor variety}. Algorithms tend to generate only one particular shape of mountainous terrain. &
	
	\textbf{Unrealistic}. Only superficially mimics the statistical distribution and shape of terrain. \\ \hline
	
		Erosion Simulations \cite{St'ava08} &
		
	\textbf{Slow}. Simulations take at least 50 and usually a few hundred iterations to stabilize, generally speaking. Takes a few minutes in total. &
	
	\textbf{Poor controllability}. The running simulation may generate or remove features against the designer's wishes. Designer is largely limited to tweaking global variables, and/or modifying the setup if the simulation runs interactively.
	\textbf{Generally high usability}, as the algorithm parameters reflect real life physical parameters. However, the parameters of some algorithms may have an indirect if not obscure effect on the final terrain. &
	
	\textbf{High independence}. Only requires another algorithm or designer to provide initial terrain to run simulations on, and the initial terrain can be very simple or artificial. &
		
	\textbf{Generally poor variety}. The main source of variety is from the initial terrain, which may or may not be varied. The terrain is then run through generally the same global processes every time. &
	
	\textbf{Generally very realistic}, as they mimic actual physical processes. The conventional hydraulic erosion algorithms such as \cite{Musgrave89,Mei07,St'ava08} simulate water as it acts in real time (i.e. time steps of seconds), which is unrealistic for geological time scales. \\ \hline
	
		Machine Learning Algorithms \cite{Guerin17} &
	
	\textbf{Very fast}. Milliseconds, once the algorithm has been trained. &
	
	\textbf{High interactivity}. Designer is generally able to specify placement and distribution of terrain features, which the algorithm then fits with good precision. &
	
	\textbf{Very poor independence}. A large database of realistic terrain samples has to be acquired to train the algorithm. One or more data scientists have to engineer and refine the algorithm so that it generates good output. The algorithm requires a terrain designer at runtime to specify placement of terrain features. &
	
	\textbf{Generally very varied}. Depends on training data and designer specifications. &
	
	\textbf{Generally very realistic}. Depends on training data and competency of the machine learning algorithm. \\ \hline
	
		Graph-based Erosion \cite{Cordonnier16} &
		
	\textbf{Slow}. Takes about 100-300 iterations to stabilise. Takes up to 3 minutes in total. &
	
	\textbf{Medium interactivity}. Designer is limited to specifying the uplift map, which indirectly affects the distribution and shape of mountains. The uplift map is generally intuitive, however some unintuitive creativity is needed in order to manipulate the shape of the mountains, e.g. making the uplift map piecewise instead of smooth. &
	
	\textbf{High independence}. The uplift map may be fed with output from another fast terrain generator, such as fractal noise, which can be very simple or artificial. Uplift and other parameters may be tweaked mid-simulation. &
		
	\textbf{Good variety}. Uplift is a locally varying parameter of the simulation. &
	
	\textbf{Very realistic}. Uplift and fluvial erosion mimic actual physical processes. \\ \hline
	
		Our algorithm &
		
	\textbf{Medium-Slow}. Takes about 50-150 iterations to achieve realistic terrain, depending on how isolated the most inland regions are, although the simulation can go on forever without stabilizing. Takes up to 2 minutes in total. &
	
	\textbf{High interactivity}. Designer is able to specify the constraint, constraint strengths, and moisture maps. Each provides high and intuitive control over the distribution and shape of mountains, details, and erosion. &
	
	\textbf{High independence}. The constraint map may be fed with output from another fast terrain generator, such as fractal noise, and good default values have been recommended for the rest, generating competitive results. Constraints, moisture and other parameters may be tweaked mid-simulation. &
		
	\textbf{Good variety}. Constraint and moisture are locally varying parameters of simulation. The height value and gradient constraints preserve variety at the large and small scales respectively. &
	
	\textbf{Very realistic}. Fluvial erosion mimics the actual physical process, while gorge carving generates visually similar gorges to the real world. \\ \hline
\end{tabular*}
\end{scriptsize}
\end{center}
\end{table} 
\end{landscape}
\section*{\uppercase{Acknowledgements}}
	This work is supported by the Singapore Ministry of Education Academic Research grant T1 251RES1812, “Dynamic Hybrid Real-time Rendering with Hardware Accelerated Ray-tracing and Rasterization for Interactive Applications”. 
	
\bibliographystyle{apalike}
{\small
\bibliography{pcgNew}}

%
%

\end{document}